\documentclass[aps,preprint,nofootinbib,%
superscriptaddress,amsmath,amssymb,showpacs]{revtex4}
\usepackage{epsfig}
\usepackage{bm}

\begin{document}
\preprint{}
\newcommand{\U}{\mathrm{U}}
\newcommand{\Th}{\mathrm{Th}}
\title{KamLAND  results and the radiogenic terrestrial heat}
\author{Gianni Fiorentini}
\email{fiorenti@fe.infn.it} \affiliation{Dipartimento di Fisica,
Universit\`a di Ferrara, I-44100 Ferrara, Italy}
\affiliation{Istituto Nazionale di Fisica Nucleare, Sezione di
Ferrara, I-44100 Ferrara, Italy}
\author{Marcello Lissia}
\email{marcello.lissia@ca.infn.it} \affiliation{Istituto Nazionale
di Fisica Nucleare, Sezione di Cagliari,
             I-09042 Monserrato (CA), Italy}
\affiliation{Dipartimento di Fisica, Universit\`a di Cagliari,
             I-09042 Monserrato (CA), Italy}
\author{Fabio Mantovani}
\email{mantovani@fe.infn.it} \affiliation{Dipartimento di Scienze
della Terra, Universit\`a di Siena, I-53100 Siena, Italy}
\affiliation{Centro di GeoTecnologie CGT, I-52027 San Giovanni
Valdarno, Italy} \affiliation{Istituto Nazionale di Fisica
Nucleare, Sezione di Ferrara, I-44100 Ferrara, Italy}
\author{Barbara Ricci}
\email{ricci@fe.infn.it} \affiliation{Dipartimento di Fisica,
Universit\`a di Ferrara, I-44100 Ferrara, Italy}
\affiliation{Istituto Nazionale di Fisica Nucleare, Sezione di
Ferrara, I-44100 Ferrara, Italy}
%
\date{September 2, 2005}

\begin{abstract}
We find that recent results from  the KamLAND collaboration on
geologically produced antineutrinos, N(U+Th) = 28$^{+16}_{-15}$
events, correspond to a radiogenic heat production from Uranium
and Thorium decay chains H(U+Th) = 38$^{+35}_{-33}$ TW. The 99\%
confidence limit on the geo-neutrino signal translates into the
upper bound  H(U+Th) $<162$ TW, which is much weaker than that
claimed by KamLAND, H(U+Th) $< 60$ TW, based on a too narrow class
of geological models. We also performed an analysis of KamLAND
data including  recent high precision measurements of the
$^{13}$C$(\alpha,n)^{16}$O cross section. The result, N(U+Th) =
31$^{+14}_{-13}$, corroborates the evidence ($\simeq 2.5\sigma$)
for  geo-neutrinos in KamLAND data.
\end{abstract}
 \maketitle
\section{Introduction}

Geologically produced antineutrinos (geo-neutrinos) were
introduced by Eder \cite{Eder1966} in the sixties and Marx
\cite{Marx1969} soon realized their relevance. In the eighties
Krauss et al. discussed their potential as probes of the Earth's
interior in an extensive publication \cite{Krauss1984}. In the
nineties the first paper on a geophysical journal was published by
Kobayashi et al. \cite{Kobayashi1991}. In 1998, Raghavan et al.
\cite{Raghavan1998} and Rotschild et al. \cite{Rotschild1998}
pointed out that KamLAND and Borexino should be capable of
geo-neutrino detection.

The potential of geo-neutrinos for providing information on the
energetics and composition of the Earth has been discussed in
Refs.~\cite{Fiorentini2003,Mantovani2004} and more recently in
Ref.~\cite{Fiorentini2005}.  The KamLAND collaboration has just
published \cite{Araki2005} their experimental results, claiming some
28 geo-neutrino events from Uranium and Thorium decay chains in a
two-year exposure. This important step shows that the technique for
exploiting geo-neutrinos  in the investigation of the Earth's
interior is now available. In order to understand where to go with
geo-neutrinos, we have to know where we stand, in the light of the
available data. In this spirit, the aim of this letter is to discuss
the implication of the KamLAND result on the contribution of Uranium
and  Thorium decay chains to the terrestrial heat.

\section{The geo-neutrino signal and the radiogenic terrestrial heat}
For a given value of, {\em e.g.},  the Uranium mass in the Earth,
$m(\U)$, the contributed heat production rate from the Uranium
decay chain is uniquely determined, H(U) = 0.95~TW $\times m(\U) /
(10^{16} \mathrm{kg})$, whereas the flux and signal of
geo-neutrinos depend on the detector location and on the Uranium
distribution inside the Earth. The connection between the signal
of geo-neutrinos from the Uranium decay chain, the mass of Uranium
in the Earth  and the heat production rate from that element was
found in Ref.~\cite{Fiorentini2005},  by using global mass balance
together with a detailed geochemical and geophysical study of the
region surrounding Kamioka.

We remark that the mass of Uranium in the crust, $m_c(\U)$, is
rather well constrained by geological data, in the interval $(3
\div 4)\times 10^{16}$~kg.  The main uncertainty is the amount of
Uranium (and Thorium) in the mantle. Geo-neutrinos should provide
us with this information.

For a total uranium mass $m(\U)$ in the Earth, the maximal and
minimal signals  can be derived by using a proximity argument: the
maximal (minimal) signal is obtained by placing the sources as
close (as far) as possible to (from) the detector, consistently
with geochemical and geophysical constraints. The maximal signal
is thus obtained  for an Uranium rich crust, $m_c(\U)=4\times
10^{16}$ kg, and distributing uniformly in the mantle the rest of
Uranium, $m(\U)-m_c(\U)$. The minimal signal corresponds to an
Uranium poor crust, $m_c(\U)=3\times 10^{16}$ kg, and distributing
the rest near the bottom of the lower mantle. These maximal and
minimal signals provide the borders of Fig.~5 of
Ref.~\cite{Fiorentini2005}, where the interested reader can find
more details.

\begin{figure}
\begin{center}
  \epsfig{figure=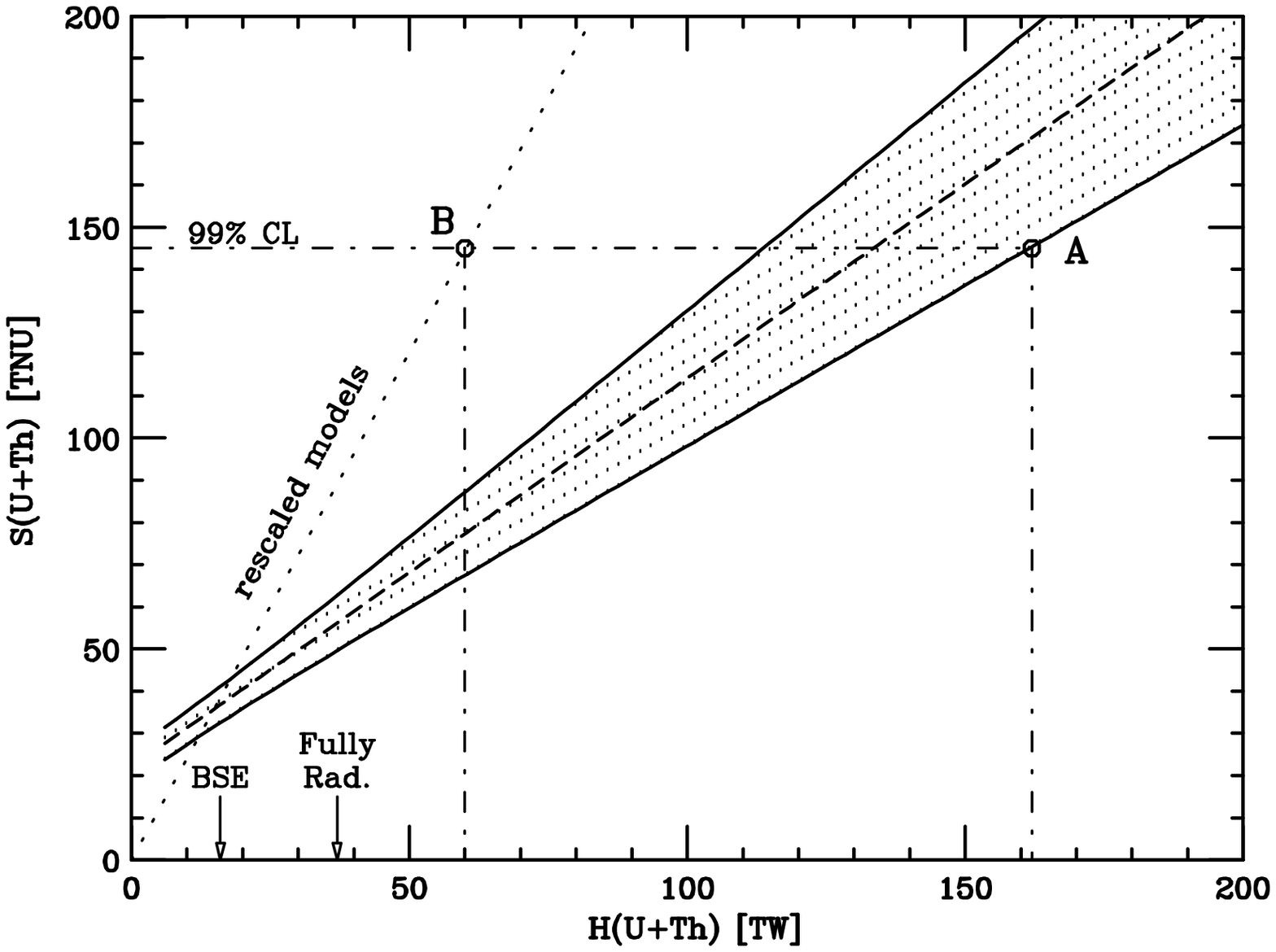,width=0.8\hsize}
\end{center}
\caption[b]{ The combined signal from  Uranium and Thorium
geo-neutrinos and the radiogenic heat production rate.

\ \ The shaded area denotes the region allowed by geochemical and
geophysical constraints. The dashed  median line represents our
best estimate for the relationship between signal and  power.

\ \ The dotted line denotes the ``rescaled models'' of Eq.
\ref{eq_retta_kamland}, used in \cite{Araki2005}. Note that most
of these models are outside the allowed area.

\ \ One TNU corresponds to 10$^{-32}$  $\bar{\nu_e}$ captures per
target proton per year. \label{fig:1SvsH}}
\end{figure}

We have extended the same calculations to Thorium, assuming a
global chondritic Uranium-to-Thorium mass ratio,
$m(\Th)/m(\U)=3.9\pm 0.1$, so that we can now connect the combined
signal of geo-neutrinos from Uranium and Thorium progenies,
S(U+Th), with the radiogenic heat production rate from these
elements, H(U+Th), see Fig.~\ref{fig:1SvsH}.

The geo-neutrino signal is expressed in Terrestrial Neutrino
Units, one TNU corresponding to $10^{-32}\bar{\nu_e}$ captures per
target proton per year.

The  allowed band in Fig.~\ref{fig:1SvsH} is estimated by
considering {\em rather extreme} \footnote{We are neglecting here
the possibility that some Uranium or Thorium is hidden in the
core. This possibility, which has been advanced by some authors
(see, e.g., Herndon ~\cite{Herndon1996} and Hofmeister and Criss
\cite{Hofmeister2005}) would imply an even smaller signal-to-power
ratio.} models for the distributions of radioactive elements,
chosen so as to maximize or minimize the signal for a given heat
production rate~\cite{Fiorentini2005}.

We also remark that, in comparison with the experimental error,the
width of the band is so narrow that we can limit the discussion to
the median line of the allowed band in Fig.~\ref{fig:1SvsH}, which
represents our best estimate for the relationship between signal
and power.

For the Bulk Silicate Earth (BSE) model, H(U+Th) = 16 TW, our
prediction for Kamioka is centered at 37 TNU.

By assuming that Uranium and  Potassium in the Earth are in the
ratio 1/10,000 and that there is no Potassium in the core, the
total radiogenic power is H(U+Th+K) = 1.18 H(U+Th). With these
assumptions, a maximal and fully radiogenic heat production rate,
H(U+Th+K) = 44 TW, corresponds to H(U+Th) = 37 TW, which gives a
signal S(U+Th) $\approx 56$ TNU.

The KamLAND collaboration has reported \cite{Araki2005} data from
an exposure of N$_p=(0.346 \pm 0.017)\times 10^{32}$ free protons
over a time  $T=749$ days with a detection  efficiency $\epsilon=
69\%$: the effective exposure is thus  E$_\mathrm{eff}$=N$_p
\times T \times \epsilon=(0.487 \pm 0.025) \times 10^{32}$ protons
$\cdot$ yr. In the energy region where geo-neutrinos are expected,
there are C = 152 counts, implying a statistical fluctuation of
$\pm$12.5. Of these counts, a number R $= 80.4\pm7.2$ are
attributed to reactor events, based on an independent analysis of
higher energy data. Fake geo-neutrino events, originating from
$^{13}$C$(\alpha,n)^{16}$O reactions following the  alpha decay of
contaminant  $^{210}$Po, are estimated to be F $= 42\pm 11$, where
the error is due to  a 20\% uncertainty on the
$^{13}$C($\alpha$,n)$^{16}$O cross section and a 14\% uncertainty
on the number  of  $^{210}$Po  decays in the detector. Other minor
backgrounds account  for B $=4.6\pm 0.2$ events. The number of
geo-neutrino events is estimated  by subtraction, N(U+Th) = C - R
- F - B, with an uncertainty obtained by combining the independent
errors: N(U+Th) $= 25^{+19}_{-18}$. The geo-neutrino signal is
thus S(U+Th) = N(U+Th) / E$_\textrm{eff}= 51^{+39}_{-36}$ TNU.
From the median line in Fig.~\ref{fig:1SvsH} one finds
\begin{equation}
    \mbox{H(U+Th)} =    31  ^{+43}_{-31} \mbox{ TW       \qquad  (rate
    only)} \quad .
\end{equation}
This ``rate only'' study  has been improved in
Ref.~\cite{Araki2005} by exploiting  the shape of the spectrum. A
likelihood analysis of the unbinned spectrum yields N(U+Th) =
$28^{+16}_{-15}$, see Fig.~4b of Ref.~\cite{Araki2005}. This
implies S(U+Th) $=57^{+33}_{-31}$ TNU and
\begin{equation}
    \mbox{H(U+Th)} =    38  ^{+35}_{-33}
    \mbox{ TW      \qquad    (rate + spectrum)}\quad .
\end{equation}
The best fit value is close to the maximal and fully radiogenic model, however the BSE is
within $1\sigma$.

By using the median line in Fig.~\ref{fig:1SvsH}, the 99\%
confidence limit on the signal (145 TNU) corresponds to 133 TW. If
we include the uncertainty band of the theoretical models, we find
an upper bound of 162 TW, see point A in Fig.~\ref{fig:1SvsH}.
This point corresponds to a model with a total Uranim mass $m(\U)
= 80 \times 10^{16}$ kg, an uranium poor crust, $m_c(\U)= 3 \times
10^{16}$ kg, the rest of the Uranium being placed at the bottom of
the mantle, and global chondritic Thorium-to-Uranium ratio.

This 162 TW upper bound is much higher than the 60 TW upper bound
claimed in Ref.~\cite{Araki2005}, which was obtained by using a
family of geological models which are {\em too narrow}  and are
also {\em incompatible} with well-known geochemical and
geophysical data.

In fact, the authors of Ref.~\cite{Araki2005} start with a
reference BSE model derived from Ref.~\cite{Sanshiro2005}: the
total Uranium mass is $m_\textrm{BSE}= 8\times 10^{16}$ kg,
roughly half in the crust and the rest in the mantle,  and the
abundance ratio is Th/U = 3.9. This model  corresponds to
H$_\textrm{BSE}(\U+\Th)=16$ TW and predicts a signal of 38.5 TNU,
very close to our prediction for BSE. The signal {\em is assumed}
to scale with  the total mass of U+Th, so that heat production and
signal are also proportional:
\begin{equation}
\label{eq_retta_kamland}
  \textrm{S(U+Th)} = 38.5 \textrm{ TNU } \times  \textrm{H(U+Th)}
  / (16 \textrm{ TW}) \,.
\end{equation}
In this way, the 99\% upper limit on the signal, 145 TNU, is
translated into 60 TW \cite{Araki2005}, see point B in
Fig.~\ref{fig:1SvsH}.

This scaling assumption, however,  produces a too limited series
of models. The points in the shaded area of Fig.~\ref{fig:1SvsH}
correspond to all models which are compatible with available
geochemical and geophysical data~\footnote{We note that actually
models with H(U+Th)$>$ 37 TW are essentially unrealistic; this
portion of the graph is included just for discussing KamLAND
results.}: most of these models cannot be obtained by
Eq.~\ref{eq_retta_kamland} and predict, for a given signal, a
larger power than Eq. \ref{eq_retta_kamland}, which therefore
cannot be used to derive an upper bound on  the radiogenic power
production.

Furthermore, Eq.~\ref{eq_retta_kamland} implies that Uranium in
the crust (and in the mantle) scales linearly with the total
Uranium mass. This becomes incompatible with the geochemical data
on the crust ($m_c(\U) < 4 \times 10^{16}$ kg) already  for total
masses slightly above the BSE estimate, i.e., for models where
H(U+Th)$ >20$ TW.  For example, in the model yielding 60 TW (point
B in Fig.~\ref{fig:1SvsH}) the crust should contains about
$13\times 10^{16}$ kg of Uranium, four times more than the largest
geochemical estimate. This inconsistency is clearly seen in
Fig.~\ref{fig:1SvsH}, which shows that the family of models
labeled as ``rescaled models'' lies essentially in the
geo-chemically excluded region.

We remark that the  bound  H(U+Th)$<$162 TW  which we have
extracted from KamLAND data does not add any significant
information on Earth's interior, since anything exceeding H(U+Th)=
37 TW  (i.e. H(U+Th+K)=44 TW)  is unrealistic.  The upper limit
simply reflects the large uncertainty in this pioneering
experiment.

On the other hand,  what is important for deciding the potential
of future experiments is the relationship between geo-neutrino
signal and heat production in the physically interesting region,
H(U+Th)$ \leq 37$ TW.  The basic parameter is the slope, dS/dH,
which expresses how the experimental error translates into an
uncertainty on the deduced heat production. For our models we find
from Fig.~\ref{fig:1SvsH} dS/dH $\simeq 1$ TNU/TW. Discrimination
between BSE and fully radiogenic models, which requires a
precision $\Delta$H~$ \sim 7$~TW, requires thus an experiment with
an accuracy $\Delta$S~$ \sim 7$~TNU.  The ``rescaled models'' of
ref.~\cite{Araki2005}, yielding dS/dH$\simeq 2.4$ TNU/TW, might
misleadingly suggest that the same goal can be achieved for
$\Delta$S=17 TNU.

\section{The geo-neutrino signal and the $^{13}$C$(\alpha,n)^{16}$O  cross section. }

As already remarked, a  major uncertainty for extracting the
geo-neutrino signal originates from the $^{13}$C$(\alpha,n)^{16}$O
cross section\footnote{In fact, the claim of 9 geo-neutrino events
in Ref.~\cite{Eguchi2003} should be dismissed: more than half of
these events are to be considered as fake signal, produced from
$^{13}$C($\alpha,n)^{16}$O reaction.}.
The values used in Ref.~\cite{Araki2005} are taken from the JENDL
\cite{Jendl2005} compilation, which provides an R-matrix fit of
relatively old data. A 20\% overall uncertainty has been adopted
in \cite{Araki2005}, corresponding to the accuracy claimed in the
original experimental papers (see, e.g., Ref.~\cite{Bair1973}).

\begin{figure}
\begin{center}
  \epsfig{figure=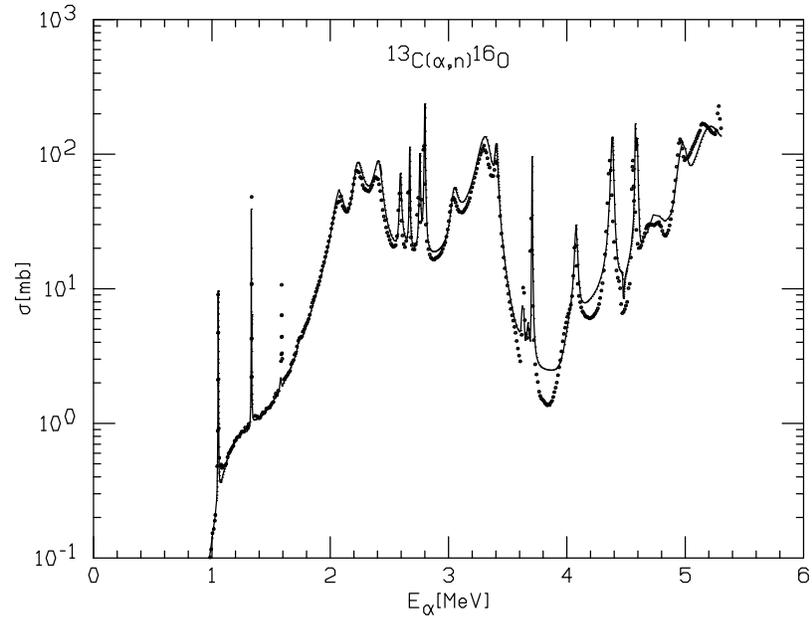,width=0.5\hsize,angle=90}
\end{center}
\caption[b]{ Cross section of $^{13}$C($\alpha$,n)$^{16}$O. The
solid line corresponds to the JENDL compilation, dots are the
experimental points from  Ref.~\cite{Harissopulos2005}.
\label{fig:2xsectC13}}
\end{figure}

Recently a series of high precision measurements for this cross
section has been performed \cite{Harissopulos2005}.  In the
relevant energy range $(1\div 5.3)$ MeV, the absolute
normalization has been determined within a 4\% accuracy. The
measured values are generally  in very good agreement with those
recommended  in JENDL, see Fig.~\ref{fig:2xsectC13}; however,  we
find that the neutron yield per alpha particle is 5\% smaller.
It follows that the number of fake neutrinos is lower, F = $40\pm
5.8$, and  geo-neutrino events obviously increase\footnote{Indeed
Ref.~\cite{Araki2005} mentions that an alternative analysis
including the time structure of the scintillation light from
different particles produced a slightly larger geo-neutrino
signal, which is consistent with the result presented here.}.

The ``rate only'' analysis gives now $27^{ +16}_{-15}$
geo-neutrino events, corresponding to S(U+Th) $= 55^{ +33}_{-31}$
TNU. From the median line of Fig.~\ref{fig:1SvsH}, the radiogenic
power is now:
\begin{equation}
\mbox{H(U+Th)} = 36^{+35}_{-33} \mbox{ TW \qquad (rate + new
$^{13}$C$(\alpha,n)^{16}$O) \, . }
\end{equation}

We also performed an analysis\footnote{A complete analysis
requires several details (the un-binned spectrum, the energy
dependence of the detection efficiency, \ldots) which are not
available to us. Just for a comparison, the binned spectrum
analysis using the JENDL cross sections with 20\% uncertainty
gives us N(U+Th)=28.5$^{+15}_{-14}$, in agreement with
\cite{Araki2005}.} of the binned spectrum reported in Fig. 3 of
Ref.~\cite{Araki2005}.
This analysis gives N(U+Th) $ =31^{+14}_{-13}$  counts,
corresponding to S(U+Th) $= 63^{+28}_{-25}$ TNU  and thus:
\begin{equation}
\mbox{H(U+Th)} = 44  ^{+31}_{-27} \mbox{ TW \qquad (rate +
spectrum + new $^{13}$C($\alpha$,n)$^{16}$O)\, . }
\end{equation}

\section{Concluding Remarks}
In  summary, the new data  on $^{13}$C$(\alpha,n)^{16}$O
corroborate the evidence for geo-neutrinos in KamLAND data, which
becomes near to 2.5$\sigma$.

On the other hand, the determination of radiogenic heat power from
geo-neutrino measurements is still  affected by a 70\%
uncertainty. The best fit of H(U+Th) is close to the prediction of
a maximal and fully radiogenic model, however  the BSE prediction
is within 1$\sigma$ from it.

With more statistics KamLAND should be capable of providing a
three sigma evidence of geo-neutrinos, but discrimination  between
BSE and fully radiogenic models definitely requires new detectors,
with class and size similar to that of KamLAND, far away from
nuclear power plants.

\section*{Acknowledgments}
We are grateful to C. Rolfs and his group for useful discussions
and for allowing us to use their results.

We thank for their useful comments A.~Bottino, E.~Lisi,
W.~F.~McDonough, and R.~Raghavan. We appreciated the suggestions
of the anonymous referee.

\end{document}